\documentclass[runningheads,citeauthoryear]{apinv}
\usepackage{epsfig,cite,graphics}
\usepackage{marvosym} 
\usepackage[utf8]{inputenc}

\begin{document}

\title{Differential and integral scaling laws of the mass density of molecular clouds}
\titlerunning{Scaling laws in molecular clouds}
\author{Sava Donkov\inst{1}, Valentin Kopchev\inst{2}, Ivan Zh. Stefanov\inst{2}, Todor V. Veltchev\inst{3,4}}
\authorrunning{S. Donkov et al}
\tocauthor{Sava Donkov, V. Kopchev, Ivan Zh. Stefanov, Todor V. Veltchev} 
\institute{Institute of Astronomy and NAO, Bulgarian Academy of Sciences, 72 Tzarigradsko Chausee Blvd., 1784 Sofia, Bulgaria
	\and Department of Applied Physics, Faculty of Applied Mathematics, Technical University-Sofia, 8 Kliment Ohridski Blvd., Sofia 1000, Bulgaria
	\and Faculty of Physics, University of Sofia, 5 James Bourchier Blvd., 1164 Sofia, Bulgaria
	\and Universit\"at Heidelberg, Zentrum f\"ur Astronomie, Institut f\"ur Theoretische Astrophysik, Albert-Ueberle-Str. 2, 69120 Heidelberg, Germany    \newline
	\email{sddonkov@astro.bas.bg}    }
\papertype{Submitted on xx.xx.xxxx; Accepted on xx.xx.xxxx}	
\maketitle

\begin{abstract}
	
In this work we investigate the mass density - size scaling law in molecular clouds (MCs). This relation reflects the fractal nature of MCs and plays a key role for understanding the physics, structure and evolution of these objects. We make use of the notion "ensemble of MCs", introduced in our previous work (Donkov, Veltchev \& Klessen, 2017), in which all MCs with the same probability density function (PDF) of mass density and effective size are represented by an abstract spherical cloud with the same PDF and size. In this spirit, the model is built on the base of abstract scales of the clouds' substructures (which are simply the radii of the spherical object). We consider two forms of the mass density - size scaling law: differential and integral, which in turn reveal the local and the global fractal clouds' structure. Both scaling functions are characterized by their scaling exponents, which can be explicitly expressed by the PDF of mass density, in the general case. Moreover, we derive a first order linear differential equation connecting the two scaling exponents and obtain its exact solution. As examples we apply this abstract construction to two PDFs: the so called power-law tail and the log-normal. Both have great importance for MC structure and evolution, as the latter corresponds to the earlier stages of clouds' evolution, when supersonic turbulence dominates the physical processes, while the former describes the latest stages of evolution, when star-formation takes place. The obtained results for the scaling exponents in both examples are qualitatively and numerically consistent with respective observations and simulations of MCs.

\end{abstract}

\section*{Introduction}
\label{Introduction}

Molecular clouds (MCs) are places of star-formation (e.g. Semkov, 2023). This is valid not only for our Galaxy, but also for a large number of nearby and distant galaxies in the Universe, whose star-formation processes have been studied. So the great importance of MCs, for galaxy structure and evolution, is undoubted (Hennebelle \& Falgarone, 2012; Klessen \& Glover, 2016; Ballesteros-Paredes et al., 2020). MCs are gaseous structures containing mostly molecular hydrogen and,  in our Galaxy, a few percent of heavier species and one percent of dust particles (Hennebelle \& Falgarone, 2012; Klessen \& Glover, 2016). Their temperatures reside in a narrow range $T\sim 10 - 30 ~ {\rm K}$ (Ferriere, 2001). Their physics is governed by gravity, (supersonic) turbulence, accretion of gas from the surrounding medium and from larger to smaller substructures in the cloud, nearly isothermal equation of thermodynamic state and magnetic fields (Hennebelle \& Falgarone, 2012; Klessen \& Glover, 2016; V\'azquez-Semadeni et al., 2019; Ballesteros-Paredes et al., 2020). The feedback of new-born stars and supernovae, which eventually disrupt the parental cloud, complete the picture.

Molecular clouds display fractal structure (see e.g. Elmegreen \& Scalo, 2004; Federrath et al., 2010; Hennebelle \& Falgarone, 2012; Klessen \& Glover, 2016) on spatial scales spanning several orders of magnitude: 0.001 pc $\leq l \leq$ 100 pc. This interval is thought to be the inertial range of turbulence (Elmegreen, 1997; Elmegreen \& Scalo, 2004; Hennebelle \& Falgarone, 2012; Klessen \& Glover, 2016), too. Additionally, the density contrast between large and small substructures spans several orders of magnitude. At scales $l \sim$ 100 pc the number density $n$ is about $10^2$ cm$^{-3}$, while at scales approaching those of pre-stellar cores ($l \leq$ 0.1 pc), $n$ is of order $10^5$ cm$^{-3}$ and more. If we look at denser gas, we arrive at scales of pre- and proto-stellar cores, and densities greater than $10^5$ cm$^{-3}$.

As a natural outcome of the fractal structure, the study of MCs, during the last four decades has revealed several scaling laws. First Larson (1981) discovered that the number density $n$ scales as $l^{-1.1}$, where $l$ is the effective size of the cloud or cloud's substructures. This leads to a scaling law for the mass: $M\sim l^{1.9}$. He also found that turbulent velocity fluctuations $\delta v$ scale as $l^{0.39}$, which was interpreted as the non-thermal velocities in the MCs caused by turbulence. These first findings for the clouds' fractality were deepened and extended in later works, including many observational (e.g. Solomon et al., 1987; Myers \& Goodman, 1988; Heyer et al., 2009), numerical (e.g. Vazquez-Semadeni, Ballesteros-Paredes \& Rodriguez, 1997; Ballesteros-Paredes \& Mac Low, 2002; Dib, Burkert \& Hujeirat, 2004; Padoan et al., 2006; Kritsuk et al., 2007; Federrath, Klessen \& Schmidt, 2008; Federrath et al., 2010) and theoretical papers (e.g. Padoan et al., 2006; Ballesteros-Paredes, 2006; Kritsuk et al., 2007; Federrath, Klessen \& Schmidt, 2008; Federrath et al., 2010). In these works scaling exponents for density, mass and non-thermal velocity are also specified. The listed scaling laws not only reveal the fractal cloud structure, but are also tools of theoretical modelling the cloud's physics and evolution, and hence, of the process of star-formation (e.g. Padoan \& Nordlund, 2002; Veltchev, Klessen \& Clark, 2011; Donkov, Veltchev \& Klessen, 2011; Donkov, Veltchev \& Klessen, 2012; Donkov, Stanchev \& Veltchev, 2012; Veltchev, Donkov \& Klessen, 2013; Veltchev, Donkov \& Klessen, 2016).

To our knowledge, the scaling exponents in scaling laws (i.e. mass-size or density-size), revealing the fractal structure of MCs, are defined in two different ways. They can be generalized as integral (or global), when the exponent does not depend on the scale and the respective scaling law holds for a range of spatial scales (e.g. Larson, 1981; Kritsuk et al., 2007; Federrath et al., 2010; Lombardi et al., 2010; Kauffmann et al., 2010a,b) and differential (or tangential), when the exponent is defined locally, at a given scale, as a slope of the tangent to the respective (e.g. mass-size or density-size) curve (Kauffmann et al., 2010a,b). The above two forms of scaling laws reflect the fractal nature of MCs from different points of view. The integral (global) form reveal the fractality of a cloud in a averaged sense over the range of spatial scales, in which it holds, while the differential (tangential) form uncover the local (at a given scale) properties of the fractal. Both forms complement each other in attempts to describe the MCs' structure, which is a key ingredient of any theory of star formation. Hence a good understanding of scaling laws, and their different forms, is crucial in attempts to develop our knowledge. In this study our aim is to generalize and deepen the definitions for integral and differential scaling laws, and also to find a relation (if such exists) between them. According to our definition, which is more general, the integral (or global) scaling exponent may depend on the scale, unlike some previous paper.

The present paper is dedicated on study of the mass density scaling laws in a given MC's class of equivalence (or ensemble of MCs) (both notions are introduced in Donkov, Veltchev \& Klessen, 2017 and are used as synonyms). In section 1 we present the cloud's model, which is an abstract model of the probability density function of mass density (so called PDF). In this section we remind the notion of MC's class of equivalence in terms of abstract scales. Also, we express mass, volume and mean density of cloud's substructures through the PDF, and the normalized relation between them (Donkov, Veltchev \& Klessen, 2017). In section 2 we consider two forms of the mass density scaling law: differential (local in regard to the PDF) and integral (for a whole substructure), and shortly justify the importance of these two relations. In section 3 we establish the relationship between differential and integral scaling exponents through a linear ordinary first-order differential equation. We solve the latter and its general solution, namely, the integral scaling exponent is presented through the PDF ( before, in section 2, the differential scaling exponent is, also, presented through the PDF). Then, in section 4, we demonstrate our abstract approach by use of two examples of PDFs: power-law tail and log-normal (both are very important for cloud structure and star-formation). Finally, we present a short discussion and our conclusions in section 5.

\section*{1. The model}
\label{Sec-the model}
In this section we follow the ideas, presented in Donkov, Veltchev \& Klessen (2017), to describe the cloud structure through its PDF in terms of abstract scales. In that way we can regard a set of MCs, which have one and the same PDF and effective size, as a class of equivalence, presented by one abstract class member. The class member is built as follows.

We consider a molecular cloud with effective size ${{l}_{\rm c}}$ and probability density function of mass density $P\left( s \right)$, where $s=\ln \left( \rho /{{\left\langle \rho  \right\rangle }_{\rm c}} \right)$ is the log-density with averaged cloud density ${{\left\langle \rho  \right\rangle }_{\rm c}}={{M}_{\rm c}}/{{V}_{\rm c}}$ being the normalization (where ${{M}_{\rm c}}$ and ${{V}_{\rm c}}=4/3\pi {{l}_{\rm c}}^{3}$ are the cloud`s total mass and volume, respectively).  Let us regard a fixed cut-off log-density level $s$ , and define an abstract scale $l(s)$ through the equation:

\begin{equation}
	\label{equ-abstract scale}
	l(s)= l_{\rm c} \bigg( \int_{s}^{\infty} P(s')ds' \bigg)^{1/3}
\end{equation}
One can also consider the respective mass $M(s)$, volume $V(s)$ and mean density $\overline{\rho }\left( s \right)$, corresponding to this cut-off density level, through the following equations, obtained already in Donkov, Veltchev \& Klessen (2017):

\begin{equation}
	\label{equ-M(s)-P(s)}
	M(s)={{M}_{c}}\int\limits_{s}^{\infty }{{{e}^{s'}}P\left( s' \right)ds'}~,
\end{equation}

\begin{equation}
	\label{equ-V(s)-P(s)}
	V(s)={{V}_{c}}\int\limits_{s}^{\infty }{P(s')ds'}~,
\end{equation}
and

\begin{equation}
	\label{equ-meanrho-P(s)}
	\overline{\rho}(s)= {\left\langle \rho  \right\rangle }_{\rm c} \int\limits_{s}^{\infty }{{{e}^{s'}}P\left( s' \right)ds'}/\int\limits_{s}^{\infty }{P(s')ds'}~.
\end{equation}
Now it is easy for one to see that the following normalized relation holds:

\begin{equation}
	\label{norm-equ}
	\frac{M(s)}{M_{\rm c}} = \frac{\overline{\rho}(s)}{{\left\langle \rho  \right\rangle }_{\rm c}} \frac{V(s)}{V_{\rm c}}~.
\end{equation}

The cloud we consider throughout this paper is a spherical isotropic cloud with outer radius ${{l}_{c}}$, PDF $P\left( s \right)$ and radius, mass and volume at the cut-off level $s$ defined through the equations (1), (2), and (3), accordingly. This abstract spherical cloud is considered as a statistical object, obtained through $P\left( s \right)$. In other words, this ball demonstrates the same statistical behaviour as a real cloud if we restrict our consideration only to the mass density PDF.

In order to simplify our calculations, we introduce the following dimensionless quantities: $\lambda \left( s \right)\equiv l\left( s \right)/{{l}_{c}}$ , $\omega \left( s \right)\equiv \rho \left( s \right)/{{\left\langle \rho  \right\rangle }_{c}}$ , $\overline{\omega }(s)\equiv \overline{\rho }\left( s \right)/{{\left\langle \rho  \right\rangle }_{c}}$ , which respectively denote the dimensionless radius, density and averaged density at a given cut-off level $s$ of the ball.

\section*{2. Differential and integral mass density scaling relations}
\label{Sec-diff and int mass density scaling relations}

Let us consider the relation $\overline{\omega }(\lambda )$, which we label “mass density-size relation”, or “mass density scaling relation”. The introduction of such a function can be justified by observations (e.g. Larson, 1981; Solomon et al., 1987; Myers \& Goodman, 1988; Lombardi, Alves \& Lada, 2010; Bellesteros-Paredes, D'Alessio \& Hartmann, 2012) and numerical experiments (e.g. Kritzuk et al., 2007; Bellesteros-Paredes, D'Alessio \& Hartmann, 2012). It can also be derived theoretically if one excludes the parameter $s$ from both relations: $\overline{\omega }=\overline{\omega }\left( s \right)$ and $\lambda =\lambda \left( s \right)$. If we are interested in the local behaviour of the relation $\overline{\omega }(\lambda )$ (See e.g. Kauffmann et al., 2010a; Kauffmann et al., 2010b, where the authors have introduced a differential mass scaling relation.), then the relevant form is the "differential mass density scaling relation":

\begin{equation}
	\label{equ-diff meanrho-l relation}
	d\ln \overline{\omega }=\alpha \left( s(\lambda) \right)d\ln \lambda~.
\end{equation}
In fact, the equation (6) simply defines the derivative of the function $\overline{\omega }(\lambda )$ in logarithmic scale, hence $\alpha \left( s \right)=\alpha \left( s\left( \lambda  \right) \right)$ define the tangent to the curve $\overline{\omega }(\lambda )$ in log-scale.

Can one determine the exponent $\alpha \left( s \right)$ through the PDF $P\left( s \right)$? If the general structure of the MC is encoded in its PDF, then the answer must be positive. To obtain the explicit form of $\alpha \left( s \right)$ one can make use of equations (1) and (4) (in dimensionless form) and use the expressions for $\lambda$ and $\overline{\omega }$ in equation (6). Then, after some algebra one obtains the following:

\begin{equation}
	\label{equ-diff exponent through the PDF}
	\alpha \left( s \right)=3\frac{{{e}^{s}}\int\limits_{s}^{\infty }{P\left( s' \right)ds'}-\int\limits_{s}^{\infty }{{{e}^{s'}}P\left( s' \right)ds'}}{\int\limits_{s}^{\infty }{{{e}^{s'}}P\left( s' \right)ds'}}~,
\end{equation}
which is the wanted expression.
 
In observational and in numerical works (and sometimes in theoretical papers) the authors usually derive the relation between the averaged density $\overline{\omega }$ and the effective scale $\lambda$ for a whole molecular cloud and/or some regions (substructures) in this cloud. In view of the fractal structure of MCs, they write this relation in the form: $\overline{\omega }={{\lambda }^{a}}$, usually assuming the exponent $a$ is a constant in regard to the scale $\lambda$ (see e.g. Kauffmann et al., 2010a,b), and therefore $a=\alpha ={\rm const}$. However, a careful study of molecular cloud structure reveals that MCs are multi-fractals and the exponent $a$ depends on $\lambda$ (e.g. Lombardi, Alves \& Lada, 2010). Then the correct scaling relation must be as follows:

\begin{equation}
	\label{equ-int meanrho-l relation}
	\overline{\omega }={{\lambda }^{a\left( \lambda  \right)}}~.
\end{equation}
We denote the latter as the "integral mass density scaling relation".

\section*{3. Relationship between the differential and integral mass density scaling exponents}
\label{Sec-relation between the diff and int mass density scaling relations}

In order to attain a complete picture of the structure of a MC as described by its mass density PDF, it is worth to search for a link between the scaling exponents $a\left( \lambda  \right)$ and $\alpha \left( \lambda  \right)$. Thanks to equation (7), we can obtain $\alpha \left( \lambda \left( s \right) \right)$ for an arbitrary cut-off level $s$, hence if we can express $a$ through $\alpha$, then we can get the integral exponent for an arbitrary cut-off level $s$. This problem can be solved in the following steps. First, we take the logarithm of equation (8) and then differentiate both sides:

\begin{equation}
	\label{equ-algebra}
	d\ln \overline{\omega }=d\left[ a\left( \lambda  \right)\ln \lambda  \right]=\ln \lambda \frac{da}{d\lambda }d\lambda +ad\ln \lambda =\left[ a+\lambda \ln \lambda \frac{da}{d\lambda } \right]d\ln \lambda~.
\end{equation}
Then we substitute the left hand side of the above expression with equation (6) and after some simple algebra we get:

\begin{equation}
	\label{equ-diff equ}
	\alpha \left( \lambda  \right)=a\left( \lambda  \right)+\lambda \ln \lambda \frac{da\left( \lambda  \right)}{d\lambda }\quad \Leftrightarrow \quad \frac{da\left( \lambda  \right)}{d\lambda }+\frac{a\left( \lambda  \right)}{\lambda \ln \lambda }=\frac{\alpha \left( \lambda  \right)}{\lambda \ln \lambda }~,
\end{equation}
which is a linear ordinary first-order differential equation for the unknown function $a\left( \lambda  \right)$. Fortunately, this ODE can be solved explicitly. The solution reads:

\begin{equation}
	\label{equ-diff equ solution}
	a\left( \lambda  \right)=\left( -\frac{1}{\ln \lambda } \right)\left[ C-\int{\frac{\alpha \left( \lambda  \right)}{\lambda }d\lambda } \right]~,
\end{equation}
where the constant $C$ should be obtained from the requirement that the solution must converge at the outer edge of the cloud, i.e. at $\lambda =1$.

It should be noted that (11) has an explicit representation through the PDF, in the general case. If $\alpha \left( \lambda \left( s \right) \right)$ is determined, in case of an arbitrary PDF $P\left( s \right)$, through equation (7), then (11) must be as follows:

\begin{equation}
	\label{equ-diff equ general solution}
	a(\lambda(s)) = 3\frac{ \ln \int\limits_{s}^{\infty } {e}^{s'} {P\left( s' \right)ds'}-\ln\int\limits_{s}^{\infty }{P\left( s' \right)ds'}}{\ln\int\limits_{s}^{\infty }{P\left( s' \right)ds'}}~.
\end{equation}
This result can be easily derived if one takes the logarithm of equation (8) and then makes use of equations (4) and (1) (this can be checked through a direct substitution in (11), but one must pay attention to the change of variables from $\lambda$ to $s$, using equation (1)). The constant $C$ is obtained under the condition $s \rightarrow - \infty$, which corresponds to $\lambda \rightarrow 1$.

\subsection*{4. Examples}
\label{Sec-examples}

In this section we demonstrate our abstract considerations in two simple, but very important for the physics of MCs examples of mass density PDFs.
The first one is the case of the so called power-law tail (PL-tail) PDF. This PDF reads:

\begin{equation}
	\label{equ-PDF PL-tail}
	P\left( s \right)=\left( -q \right){{e}^{qs}}~.
\end{equation}
Such PDF appears at the later stages of a MC’s evolution, when gravity dominates over turbulence, thermodynamics and magnetic fields in the dense material of the cloud (see Klessen, 2000; Kritsuk, Norman \& Wagner, 2011; Donkov \& Stefanov, 2019). It represents the probability density function of mass density of dense gas where star-formation takes place. If one plots the PDF in logarithmic scale, then it is a straight line with negative slope $q$, which takes typical values in the range $[-3,-1]$ (e.g. Kritsuk, Norman \& Wagner, 2011; Girichidis et al., 2014). The differential and integral scaling exponents are constants:

\begin{equation}
	\label{equ-PL-tail diff and int scaling exp}
	\alpha =a=\frac{3}{q}~,
\end{equation}
so they reside in the same range $[-3,-1]$.

The second example is the so called log-normal PDF, which is simply a Gaussian of log-density $s$. Its explicit form is as follows:

\begin{equation}
	\label{equ-PDF log-normal}
	P\left( s \right)=\frac{1}{\sqrt{2\pi {{\sigma }^{2}}}}{{e}^{-\frac{1}{2}{{\left( \frac{s+{{\sigma }^{2}}/2}{\sigma } \right)}^{2}}}}~,
\end{equation}
where $\sigma$ is the dispersion, with typical values in the range $[0.5;2]$ (see e.g. Veltchev, Klessen \& Clark, 2011; Donkov, Veltchev \& Klessen, 2011). This type of PDF is representative of the early stages of MC’s evolution, when the turbulence governs the cloud’s physics (e.g. Vazques-Semadeni, 1994; Federrath et al., 2010). In Fig.1. (left) we show four log-normals for four different typical dispersions: $\sigma= 0.5, 1, 1.5, 2$ (see e.g. Federrath et al., 2010). If $\sigma$ increases, the log-normals become more expanded with lower maxima, and $s_{\rm max}=-\sigma^2/2$ (see equation (15)) shifts to the left. The latter means that when $\sigma$ increases, the density contrasts in the cloud become stronger and the probabilities for different densities have closer values. Physically this can be explained with the role of supersonic turbulence in the cloud's medium and its relation to $\sigma$, which is given by the equation: $\sigma^2 = \ln(1 + b^2 M^2)$ (Federrath, Klessen \& Schmidt, 2008), where $0.33\leq b \leq 1$ is a dimensionless parameter characterizing the ratio between compressive and solenoidal modes of the turbulent flow, and $M = \bar{v}_{\rm turb}/c_{\rm sonic}$ is the averaged sonic Mach number of the turbulence in the cloud. Hence a stronger turbulence with domination of compressive modes causes larger $\sigma$ and stronger density contrasts, while weaker turbulent flow with domination of solenoidal modes leads to smaller $\sigma$ and weaker density contrasts (see Fig.16 in Federrath et al., 2010).

\begin{figure}[!htb]
	\begin{center}
		\centering{\epsfig{file=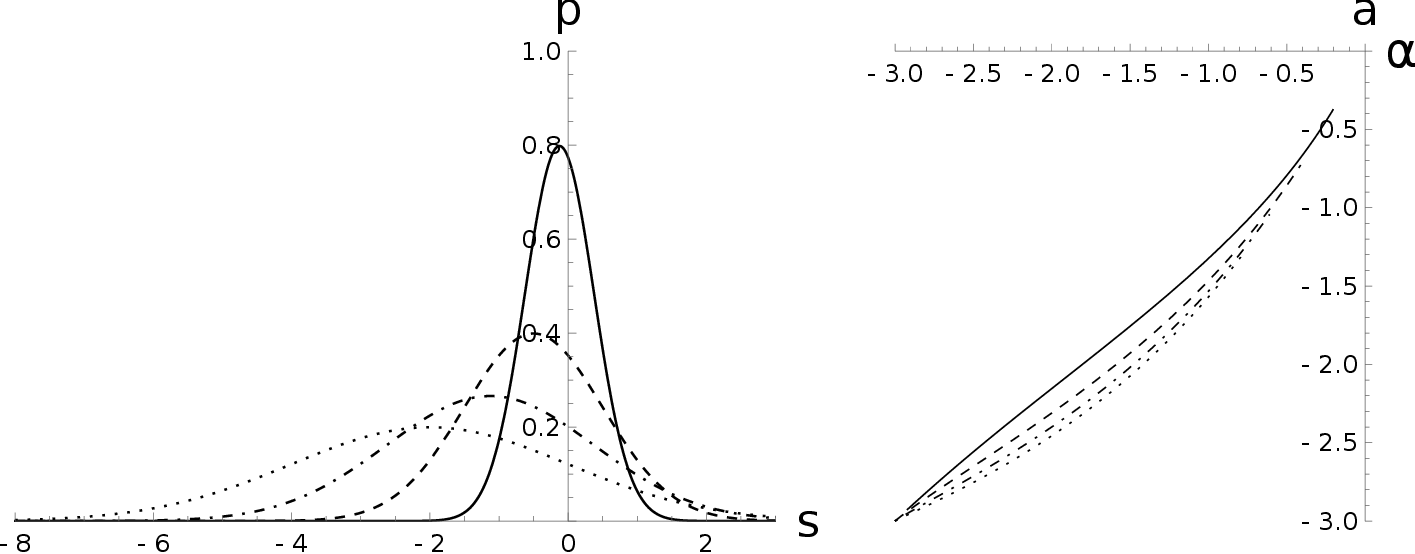, width=0.9\textwidth}}
		\caption[]{Plot of the log-normal PDFs for different dispersions: $\sigma= 0.5, 1, 1.5, 2$ (left), and the relationship between the differential and the integral scaling exponent for the same dispersions (right). If $\sigma$ increases, then the log-normals become more expanded with lower maxima (left), and the curves of the $\alpha-a$ relation decline from the identity (right).}
		\label{p_s-a_alpha}
	\end{center}
\end{figure}

 In Fig.2. we plot the differential (left column) and the integral (right column) scaling exponents, as functions of the log-density (top row) and the abstract scale (bottom row), correspondingly. Each plot shows four curves corresponding to the four different dispersions like in Fig.1., where $\sigma$ increases from the upper to the lower curves. One easily notes that $\alpha$ and $a$ are negative, and vary approximately in the range $[-3,-0.5]$, which is in qualitative agreement with Larson (1981), Lombardi, Alves \& Lada (2010), and Ballesteros-Paredes, D’Alessio \& Hartmann (2012). One can conclude also that both the differential and integral scaling exponents are increasing functions of the log-density, and decreasing functions of the abstract scale, respectively. Hence, they increase in cloud parts of increasing density, which means that the mass density scaling relations are stronger (the scaling exponents are larger in absolute value) for more diffuse parts of the cloud and therefore the density contrasts are stronger there. The latter stems from the fact that in diffuse parts of the medium the local Mach number of the turbulent flow is larger and this causes shock density fronts resulting in stronger density contrasts (see Federrath et al., 2010), reflected by the larger absolute values of scaling exponents.
 And, finally, if $\sigma$ increases, then the values of both scaling exponents decrease in both cases (log-density or abstract scale). Another interesting point is that both $\alpha$ and $a$ tend to $-3$ at the cloud's edge (when $\lambda \rightarrow 1$, and $s\rightarrow -\infty$). The latter means that $\overline{\omega } \sim \lambda^{-3}$ and therefore the normalized mass does not scale with size: $M/M_{\rm c} \sim \lambda^{a} \lambda^3 \sim \lambda^0$. This is simply the outcome of the fact that we have fixed the cloud's size $l_{\rm c}$ and mass $M_{\rm c}$ in our model. Observationally it is not trivial to determine the cloud's edge, and hence the mass scaling exponent does not tend to zero\footnote{ If one could determine the cloud's edge from observations, and uses the cloud's size to normalize the mass-size relation, then the scaling exponent of this relation must tend to zero at the cloud's edge, like in our model. } (see, e.g., Lombardi, Alves \& Lada, 2010; Ballesteros-Paredes, D’Alessio \& Hartmann, 2012). 
 
\begin{figure}[!htb]
	\begin{center}
		\centering{\epsfig{file=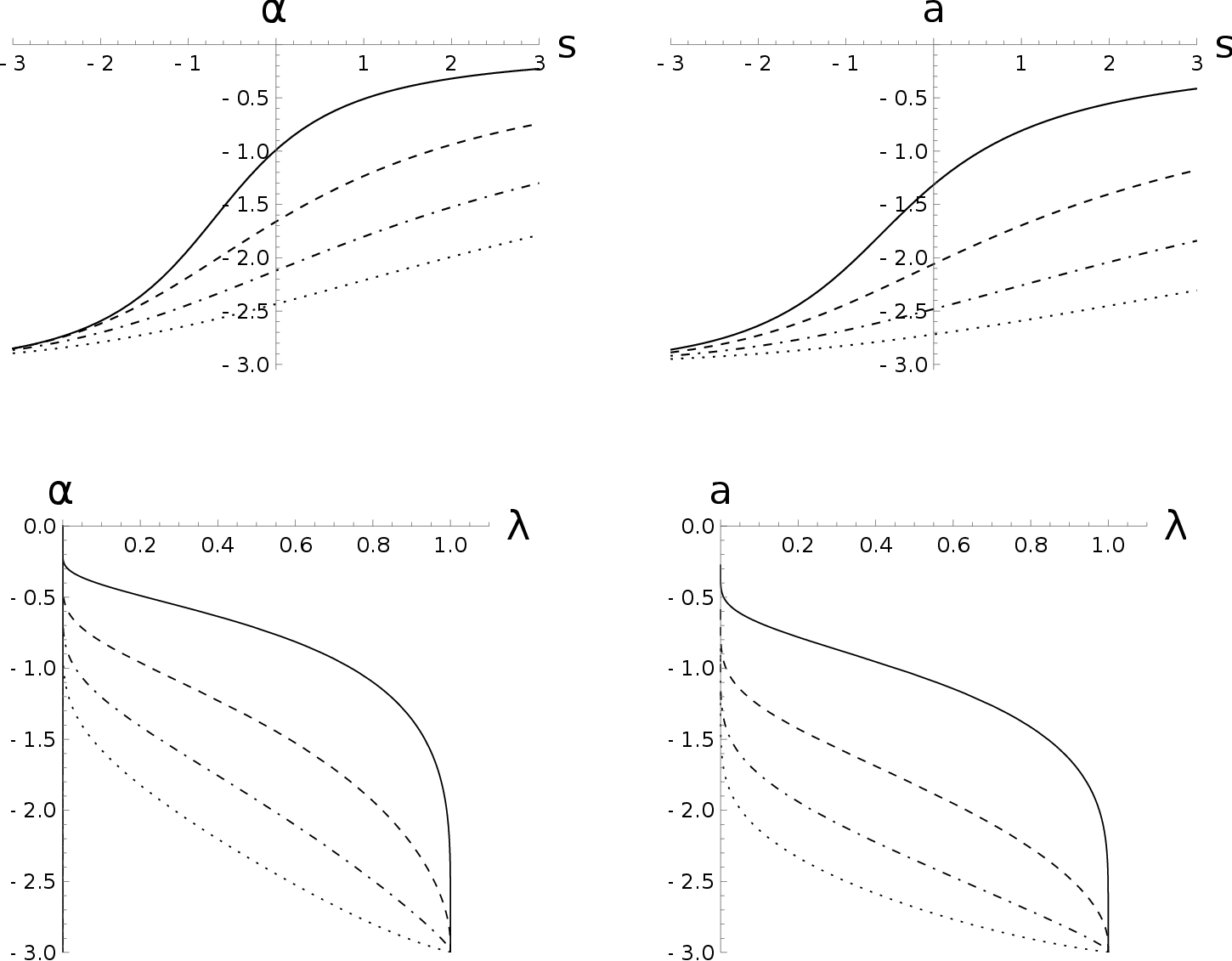, width=0.9\textwidth}}
		\caption[]{Plot of the differential (left column) and the integral (right column) scaling exponents as functions of the log-density (top row) and the abstract scale (bottom row). At every single plot there are four curves corresponding to four different dispersions like in Fig.1., as $\sigma$ increases from the upper the to lower curves.}
		\label{alpha_s_a_lambda}
	\end{center}
\end{figure}

As a final comment we note the $\alpha-a$ relation in Fig.1., right. The main feature of this plot is that the integral scaling exponent (at the ordinate) displays larger absolute values than the differential scaling exponent (at the abscissa), and this trend is stronger if $\sigma$ increases. The latter means that if the cloud demonstrates larger mass density contrasts (larger $\sigma$), the scaling of the mean density of the cloud's substructures (equation (8)) becomes stronger. This implies that clouds with larger $\sigma$ (typically these are the larger clouds, see e.g. Lombardi, Alves \& Lada, 2010; Veltchev, Klessen \& Clark, 2011; Donkov, Veltchev \& Klessen, 2011; Veltchev, Donkov \& Klessen, 2016) cause preconditions for creation of more and denser substructures, which results in a massive, stronger and faster star-formation process (Hennebelle \& Falgarone, 2012; Klessen \& Glover, 2016; V\'azquez-Semadeni et al., 2019; Ballesteros-Paredes et al., 2020).

\subsection*{5. Discussion and conclusion }
\label{Sec-discussion and conclusion}

In this short paper we intended to study the mass density scaling relation in two forms: differential and integral, describing the local and the global fractal structure of MCs. We achieved this goal in the context of our model (Donkov, Veltchev \& Klessen, 2017), where an abstract spherical cloud is representative for a whole cloud's class of equivalence, defined as an ensemble of all MCs with one and the same PDF and effective size. In terms of abstract scales, we briefly introduced differential scaling relation in equation (6) and integral scaling relation through equation (8). Both scaling exponents: $\alpha$ - differential, and $a$ - integral, are defined accordingly. The link between them is obtained in equations (10) (in differential form) and (11) (in integral form). Moreover, we derived formulae for $\alpha$ and $a$, which express them through the PDF (equations (7) and (12), respectively). Then we apply these abstract notions to two typical PDFs: the PL-tail and the log-normal, which are of great importance for the physics and evolution of MCs. Both scaling exponents (differential and integral) are presented as functions of log-density and abstract scale (the case of PL-tail is trivial in that aspect, because both exponents coincide and are constants) and their numerical values and general behaviour are in agreement with observations and numerical simulations of MCs (e.g. see reviews by Hennebelle \& Falgarone, 2012; Klessen \& Glover, 2016; V\'azquez-Semadeni et al., 2019; Ballesteros-Paredes et al., 2020).

In future work we intend to expand our analysis to the mass-size and mass-density scaling relations, as well to the link between the three respective types of scaling exponents (of the density-size, mass-size and mass-density relations). This work will also help to complete the picture of the internal connection of the latter scaling relations, which describe the fractal structure of MCs and its link to the physics governing their media.

{\it Acknowledgement:} The authors thank the anonymous referee for his/her notes, comments and suggestions, which have improved our paper. SD acknowledges funding by the Ministry of Education and Science of Bulgaria (support for the Bulgarian National Roadmap for Research Infrastructure) and for the support by German Research Foundation (DFG) under grant KL 1358/20-3. TV acknowledges the support by German Research Foundation (DFG) under grant KL 1358/20-3 and additional funding from the Ministry of Education and Science of the Republic of Bulgaria, National RI Roadmap Project D01-326/04.12.2023.

\end{document}